\documentclass{aa}  

\usepackage{graphicx}
\usepackage{txfonts}
\usepackage{lipsum}
\usepackage{subcaption}   
\usepackage{lscape} 
\usepackage{placeins}  

\begin{document}

 \title{An unidentified absorption feature at 5.11 $\mu$m on the surface of Titan and Pluto from JWST spectroscopy}
 \titlerunning{Unidentified absorption on the surface of Titan and Pluto}

   \author{B. B\'ezard\inst{1}\fnmsep\thanks{Corresponding author: bruno.bezard@obspm.fr}
        \and E. Lellouch\inst{1}
        \and M. Camarca\inst{2}
        \and J. I. Lunine\inst{2,3}
        \and E. Quirico\inst{4}
        \and C. A. Nixon\inst{5}
        \and N. A. Teanby\inst{6}
        \and P. Rannou\inst{7}
        \and S. Rodriguez\inst{8}
        \and M. Es-Sayeh\inst{8,9}
        \and S. K. Trumbo\inst{10}
        \and A. C. Souza-Feliciano\inst{11}
        \and P. Lavvas\inst{7,12}
        \and T. Bertrand\inst{1}
        \and I. Wong\inst{13}
        \and N. Pinilla-Alonso\inst{14,15}
        \and G. L. Villanueva\inst{5}
                }

   \institute{LIRA, Observatoire de Paris, Universit\'e PSL, CNRS, Sorbonne Universit\'e, Universit\'e Paris Cit\'e, 5 place Jules Janssen, 92195 Meudon, France
   \and Division of Geological and Planetary Sciences, California Institute of Technology, Pasadena, CA 91125, USA
   \and Jet Propulsion Laboratory, California Institute of Technology, Pasadena, CA 91109, USA
   \and Universit\'e Grenoble Alpes, CNRS, IPAG, 38041 Grenoble, France
   \and Solar System Exploration Division, NASA Goddard Space Flight Center, Greenbelt, MD 20771, USA
   \and School of Earth Sciences, University of Bristol, Wills Memorial Building, Queens Road, Bristol, BS8 1RJ, UK
   \and LEATP, Campus Sciences Exactes et Naturelles - BP 1039, Universit\'e de Reims Champagne-Ardenne, CNRS, 51687 Reims, France
   \and Universit\'e\ Paris Cit\'e, Institut de Physique du Globe de Paris (IPGP), CNRS, 75005 Paris, France
    \and CNRM, M\'et\'eo-France, CNRS, Universit\'e de Toulouse, Toulouse, France
    \and Department of Astronomy  \& Astrophysics, University of California, San Diego, La Jolla, CA 92093, USA
    \and Florida Space Institute, University of Central Florida, Orlando, FL 32826, USA
    \and Institut d'Astrophysique de Paris, Sorbonne Universit\'e, CNRS, UMR7095, Paris, France
    \and Space Telescope Science Institute, Baltimore, MD, USA
    \and Institute of Space Sciences and Technologies of Asturias (ICTEA), University of Oviedo, 33004 Oviedo, Spain
    \and Department of Physics, University of Oviedo, 33007 Oviedo, Spain
               }

   \date{Received June 08, 2026}

  \abstract
   {Titan possesses a thick N$_2$-CH$_4$ atmosphere that makes it difficult to study its surface spectroscopically. The chemical composition of the solid surface of Titan thus remains very uncertain.}
   {By leveraging JWST's high sensitivity and large spectral coverage, we searched for any signature from Titan's surface in the broad and less explored 5-$\mu$m atmospheric window. We also investigated the JWST spectrum of Pluto which has a thin Titan-like atmosphere.} 
   {We made selections of JWST NIRSpec and MIRI spectra around Titan's disk center and compared the NIRSpec average spectrum with a radiative transfer model including gas and haze opacity.}
    {We detected an unidentified absorption in both NIRSpec and MIRI spectra of Titan centered at 5.113 $\mu$m (1956 cm$^{-1}$) and 6-7\% deep. The width of the feature is 0.024$\pm$0.0008 $\mu$m (9.2$\pm$0.3 cm$^{-1}$) in the NIRSpec spectrum recorded on the trailing side and is possibly 25\% narrower in the MIRI spectrum of the leading side. This absorption most likely originates from the surface. We could not identify this signature among published laboratory spectra of ices relevant to Titan's atmospheric compounds but present a few plausible candidates. A 4-5\% deep absorption is also present in the MIRI spectrum of Pluto but is about 3 times broader than on Titan's trailing side.}
    {}

   \keywords{Planets and satellites: individual: Titan --
                Kuiper belt objects: individual: Pluto --
                Planets and satellites: surfaces
               }

   \maketitle
   \nolinenumbers

\section{Introduction}
Titan and Pluto possess N$_2$-dominated atmospheres containing significant amounts of CH$_4$. In both bodies, an intense photochemistry produces hydrocarbons and nitriles, eventually leading to the formation of a ubiquitous organic haze \citep[see e.g. ][]{Vuitton2025, Summers2021}. Titan and Pluto also both exhibit a complex climatic system that redistributes volatile compounds (liquid CH$_4$ on Titan; N$_2$, CH$_4$ and CO ices on Pluto) and shapes their surface morphology. Despite these similarities, Titan and Pluto differ on various aspects, most notably on their surface conditions: the $\sim$1.5-bar surface pressure on Titan exceeds by 5 orders of magnitude that on Pluto (currently around 10 $\mu$bar), while the temperature varies from 94 K on Titan to as low as 37.5 K for N$_2$ ice on Pluto. Comparing the chemical composition of the atmosphere, haze and surface of both objects can thus provide valuable information on the chemical and physical processes at work in N$_2$--CH$_4$ cold atmospheres.

The New Horizons spacecraft flying by Pluto in July 2015 provided a detailed picture of the surface diversity, showing a large basin, mountains and active glaciers with a specific geographic repartition of non-volatile (water, ammonia) and volatile ices. These ices appear reddish brown due to some complex organic material. The mission also provided detailed information on the atmospheric gas composition and on the distribution and optical properties of the haze. See \cite{Stern2021} for a detailed review of the New Horizons results. More recently, the high sensitivity of the James Webb Telescope (JWST) has enabled access to the mid-infrared range of Pluto, which had previously remained unexplored \citep{Lellouch2025}. Signatures from C$_2$H$_6$, C$_2$H$_2$, C$_2$HD, CH$_3$C$_2$H, and C$_4$H$_2$ gases were detected as well as fluorescence from CH$_4$ and CH$_3$D. In addition, the haze emission spectrum was characterized and absorption bands from CH$_4$, CH$_3$D, and C$_2$H$_4$ ices at the surface were clearly seen.

As revealed by the Cassini--Huygens mission, which explored the Saturn System between July 2004 and September 2017, the surface of Titan presents a complex morphology, including fluvial landscapes, lakes, massive dune fields, mountains and tectonic features \citep{Nixon2026}. Little is known however on the composition of the solid surface \citep{Solomonidou2025} apart from the data provided by the Huygens gas chromatograph/mass spectrometer (GCMS), which detected methane and ethane vaporizing out of the ground after landing \citep{Niemann2010}. The Huygens probe Descent Imager/Spectral Radiometer (DISR) detected a broad surface absorption centered at 1.54 $\mu$m, which may be attributed to water ice but this identification is not entirely conclusive \citep{Tomasko2005}. In the near-infrared domain 0.8--5.4 $\mu$m, the solar radiation undergoes absorption by atmospheric methane bands, the N$_2$-N$_2$ collision-induced band centered at 4.3 $\mu$m and the CO band centered at 4.7 $\mu$m. As a result, only a few narrow transparency windows (centered at 0.83, 0.94, 1.07, 1.28, 1.58, 2.0, 2.8, and 5.0 $\mu$m) give access to the albedo spectrum of the surface \citep[see Fig.\ 1 of][]{Nixon2025}. Even at these wavelengths, particularly below 3 $\mu$m, extinction caused by atmospheric haze particles significantly affects the surface signal. The surface albedo variations observed between these different windows point to compositional differences over different geological units on Titan. They have often been interpreted as mixtures of different aerosol sediments with material exposed from the underlying crust (mostly water ice) \citep{Solomonidou2025}. However, no well-defined signature of a chemical compound has yet been unambiguously detected in any of these windows despite claims for exposed H$_2$O ice \citep[e.g.][]{Griffith2003, Griffith2019} and various hydrocarbon deposits \citep{Clark2010, Singh2016}. These tentative detections are based on analyses of data from the Visible and Infrared Mapping Spectrometer (VIMS) aboard Cassini, which were limited by its low spectral resolution and low sensitivity in the 5.0-$\mu$m window. More generally, studies of surface composition based on Cassini/VIMS data and using radiative transfer and spectral mixing models are limited by the small number of atmospheric windows, the instrument's low resolution, uncertainties regarding the opacity of atmospheric methane at visible and near-infrared wavelengths, and gaps in laboratory data --- all of which, taken together, lead to ambiguities and non-uniqueness of the composition retrievals \citep[see, for example, discussion in][]{Solomonidou2024}.

Observations of Titan with the Near Infrared Spectrograph (NIRSpec) and Mid Infrared Instrument (MIRI) of the JWST have recently been presented by \cite{Nixon2025}. The first analysis of these data was focused on the atmosphere with the detection of the methyl radical (CH$_3$) and the analysis of fluorescence emission from CO and CO$_2$. 

In this letter, we present the detection of a weak and relatively narrow absorption at the surface of Titan, using JWST spectra in the atmospheric window from 4.9 to 5.4 $\mu$m, which is  the broadest and least affected by haze extinction. We also report on the detection of an absorption feature in the JWST MIRI spectrum of Pluto \citep{Lellouch2025}, centered at the same wavelength but significantly wider than on Titan.

\section{JWST Observations}

   JWST NIRSpec observations of Titan were obtained as part of the Guaranteed Time Observation (GTO) project 1251 (``Titan Climate, Composition and Clouds'', PI  Nixon). Titan was observed on 4 November 2022 across the full 0.95--5.27 $\mu$m range with a resolving power of 1500 to 3500, using the Integral Field Unit (IFU) of NIRSpec \citep{Jakobsen2022, Boker2023} . Full-range spectra over 30$\times$30 imaging elements each 0.1''$\times$0.1'' in size were thus obtained. A four-point cycling dither was used for data acquisition. The total exposure time for the 2.87--5.27 $\mu$m spectral range (Grating G395H), of specific interest here, was 902 s. The sub-observer latitude and longitude on Titan were 15.1$^{\circ}$N and 261$^{\circ}$W respectively, i.e.\ approximately centered on the trailing hemisphere of the satellite. Titan's angular diameter was 0.73 arcsec.
   
   The NIRSpec data were processed through Version 1.19.1 of the JWST calibration pipeline  \citep{Bushouse2025} with the reference context map \texttt{jwst\_1468.pmap} of the JWST
Calibration Reference Data System (CRDS). Additional steps were applied as described in Supplement B.2 of \cite{Nixon2025}.
   
   Also part of the GTO 1251 project, JWST MIRI observations were conducted in the spectral imaging
mode \citep[Medium Resolution Spectroscopy (MRS);][]{Wells2015, Argyriou2023} on 11 July 2023. The complete spectral range 4.9--27.9  $\mu$m was recorded with a resolving power of 1500 to 3500. For Channel 1A (4.90-5.74 $\mu$m), investigated here, the pixel size is 0.2''$\times$0.2'' and the exposure time was set to 854 s. As for NIRSpec observations, a four-point cycling dither was performed. Observations were targeted at Titan's leading side, with center-of-disk coordinates of latitude 7.4$^{\circ}$N and longitude 86$^{\circ}$W. Titan's angular diameter was 0.78 arcsec.
   
   MIRI data reduction was performed using the 1.20.2 version of the JWST pipeline with the CRDS Context \texttt{jwst\_1475.pmap}. Compared with the standard pipeline, the weighting was changed from \texttt{drizzle} to \texttt{EMSM} to improve the baseline and the alignment was changed from \texttt{SKYALIGN} to \texttt{IFUALIGN} to remove an interpolation/regridding step. Data from the four dither positions were combined through the pipeline to produce a single datacube.
   
   Both the NIRSpec and MIRI data were geo-referenced in the manner described in \cite{Nixon2025}, a task that primarily involves the determination of the (fractional) spaxel value of the center of Titan, and  the subsequent cube recentering and calculation of all relevant geometrical parameters  for each spaxel. For MIRI, it was noted that spaxels beyond 2" from disk center, where no flux is expected, still presented non-zero radiances of unknown origin, slowly varying with wavelength, at the typical level of $\pm$200 MJy sr$^{-1}$. This mean ``background'', negative at 5.11 $\mu$m, was spectrally subtracted from the cube; at disk center, this correction induces a decrease of the 5.11-$\mu$m band depth (discussed below) by 6 \% of its value.

   More recent observations of Titan with MIRI and NIRSpec have been obtained (GTO project 2760, PI Lunine) and will be presented in a forthcoming publication (Camarca et al. in prep). In particular, the additional NIRSpec observations were obtained at four different sub-observer longitudes, permitting the global mapping of the surface albedo spectrum in the 5-$\mu$m window. Although a preliminary map could be obtained from the single GTO 1251 NIRSpec observation presented here, we opt to include this map in the upcoming paper.
   
   On 4 May 2023, Pluto was observed with JWST MIRI as part of the JWST GO-1 program 1658 (``Pluto's climate system with JWST'', PI Lellouch). Observations were acquired with the MRS observing mode as a four-position dither over the full 4.9--27.9 $\mu$m  spectral range. Details of the acquisition settings are given in \cite{Lellouch2025}. Sub-observer latitude and longitude range were 59.6$^{\circ}$N and 353-332$^{\circ}$E respectively. Pluto's angular size was 0.095 arcsec, smaller than the MIRI pixel size; therefore the MRS spectrum is averaged over Pluto and its entire atmosphere.The data reduction process is described in \cite{Lellouch2025} and is not repeated here.
   
   As in \cite{Nixon2025} and \cite{Lellouch2025}, we converted the NIRSpec and MIRI intensities (expressed in MJy sr$^{-1}$) into I/F reflectivities using the ACE-FTS atlas of solar lines of \cite{Hase2010} combined with the solar continuum from R.L. Kurucz\footnote{\url{http://kurucz.harvard.edu/sun.html}}.
   
             \begin{figure}[ht!]
   \centering
   \includegraphics[width=\hsize]{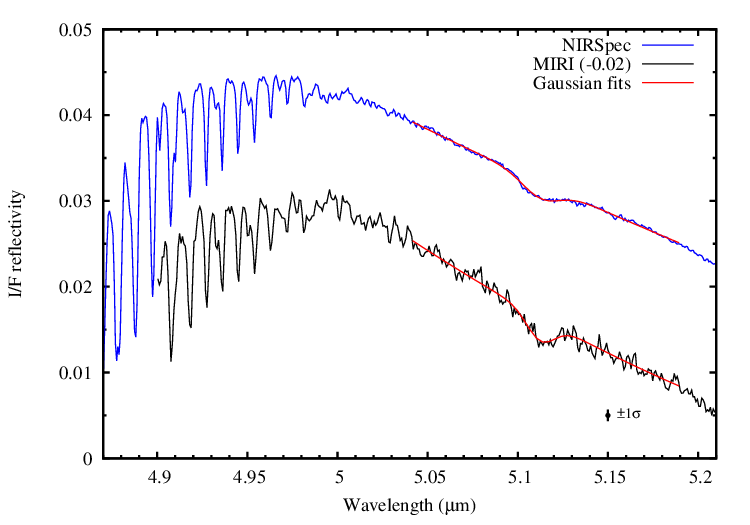}
      \caption{NIRSpec (blue) and MIRI (black) average of nadir spectra of Titan in the 5-$\mu$m atmospheric window (see text for details). The MIRI spectrum is shifted downwards by 0.02 for clarity. Red lines represent Gaussian fits of the absorption feature present at 5.11 $\mu$m. The fitting function is the sum of a second-order polynomial (three free parameters) and a Gaussian function with three additional free parameters (position, amplitude and width). The fitting interval is 5.04-5.19 $\mu$m. The $\pm$1$\sigma$ error bar for the MIRI spectrum, based on the residuals of the fit, is indicated. The $\pm$1$\sigma$ error bar of the NIRSpec spectrum, also based on the residuals of the fit, is about $\pm$0.00018 in I/F units.}
         \label{Titan1}
   \end{figure}

   \section{Results}
    \subsection{Titan} 
   
    Figure~\ref{Titan1} shows NIRSpec and MIRI spectra averaged over the four dither positions and over the 3$\times$3  (NIRSpec) and 2$\times$2 (MIRI) centermost pixels. The 5-$\mu$m atmospheric window is limited at short wavelengths by the CO (1--0) fundamental band and at long wavelengths by the far wings of the $\nu_4$ and $\nu_2$ bands of CH$_4$ (centered at 7.7 and 6.5 $\mu$m). Both spectra clearly show an absorption band centered near 5.11 $\mu$m. Note that the Cassini/VIMS instrument only marginally covered the spectral region of this feature, with reduced sensitivity and spectral sampling near the detector cutoff, preventing a reliable detection. Since this feature is detected in both MIRI and NIRSpec spectra, an instrumental artefact seems excluded. We also note that no absorption is detected in the NIRSpec spectra of Ganymede at 5.11 $\mu$m \citep{Bockelee2024}.

      \begin{figure}[ht!]
   \centering
   \includegraphics[width=\hsize]{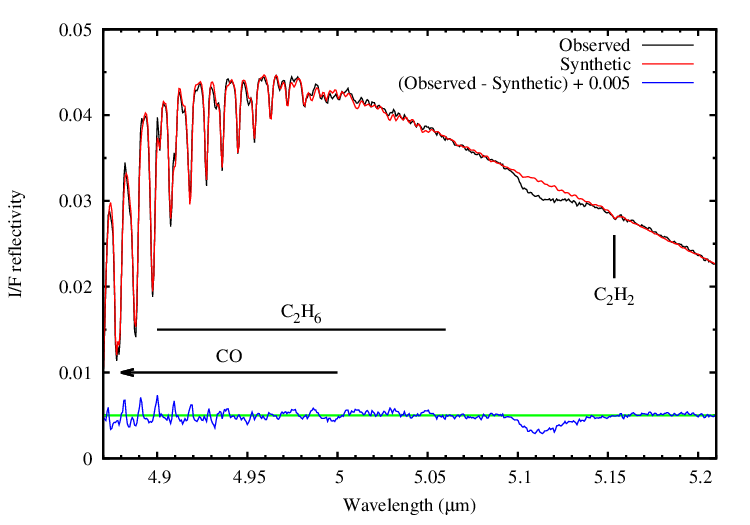}
      \caption{NIRSpec average spectrum of Titan (black) compared with a radiative transfer calculation in which the surface albedo decreases smoothly with wavelength beyond 4.9 $\mu$m (red). The blue line shows the difference between the observed and synthetic spectra, shifted by 0.005 for clarity. The 5.11-$\mu$m absorption is clearly missing in the model. The small mismatch below 4.93 $\mu$m is due to a component of the non-LTE CO emission not included in the model.}
         \label{Titan2}
   \end{figure}

    The spectral range shown here exhibits lines of $^{13}$CO and C$^{18}$O up to 5.0 $\mu$m (Fig.~\ref{Titan2}). Other weak absorption features are due to C$_2$H$_6$, visible between 4.90 and 5.06 $\mu$m, and the 2$\nu_4$+$\nu_2$ band of C$_2$H$_2$ at 5.154 $\mu$m. In Fig.~\ref{Titan2} is shown a radiative transfer calculation with the atmospheric model described in \cite{Nixon2025}. Molecular opacity from CH$_4$, CH$_3$D, CO, C$_2$H$_6$, C$_2$H$_2$  and C$_2$H$_4$ is included in this calculation. We used the haze vertical opacity profile and phase functions derived by \cite{Doose2016} from in situ Huygens measurements and assumed that the aerosol single-scattering albedo decreases linearly from 0.45 at 5.0 $\mu$m \citep[as in][]{Nixon2025} to 0.43 at 5.2 $\mu$m. The surface albedo was adjusted every 0.05 $\mu$m to best reproduce the NIRSpec spectrum and linearly interpolated in-between. The so-derived albedo decreases smoothly with wavelength from 0.060 at 4.88 $\mu$m to 0.024 at 5.2 $\mu$m. 
  Our radiative transfer calculations indicate that, beyond 4.9 $\mu$m, reflection by the surface is the dominant source of emission but haze scattering and thermal emission bring a significant contribution ($\sim$18\% at 5.1 $\mu$m). A comparison with the NIRSpec spectrum (blue line in Fig.~\ref{Titan2}) clearly shows that the 5.11-$\mu$m absorption feature mentioned above is not reproduced by the model, while very subtle atmospheric features, e.g. at 5.00-5.05  $\mu$m and C$_2$H$_2$ at 5.154  $\mu$m, are.
    
         \begin{figure}[ht!]
   \centering
   \includegraphics[width=\hsize]{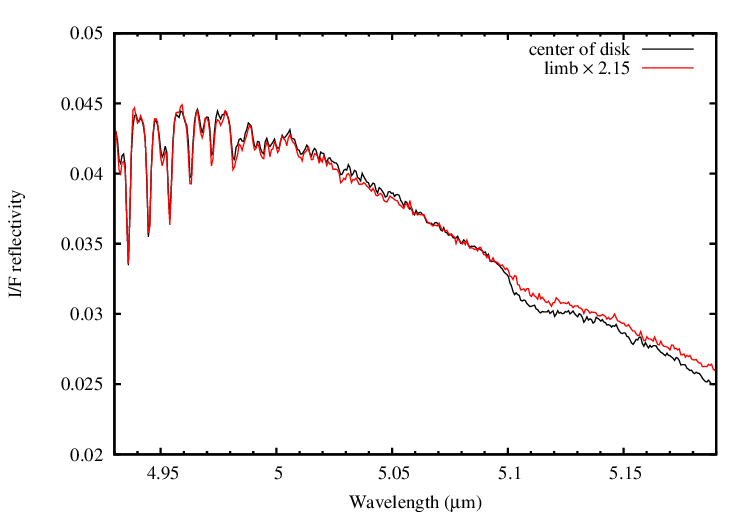}
      \caption{Average of NIRSpec spectra recorded at Titan's disk center (black) compared with an average of spectra recorded at the limb, with a field-of-view centered between 0.865 and 1.135 Titan radii from disk center (red). The limb selection has been multiplied by a factor of 2.15 to ease comparison of the relative absorption depths. The CO line-to-continuum ratios are similar in the two selections while the 5.11-$\mu$m feature is half weaker in the limb selection. As discussed in the text, this behavior suggests that this feature originates from the surface.}
         \label{Titan3}
   \end{figure}

   This structure-less feature, broader than a typical molecular Q-branch, does not resemble an atmospheric absorption. It cannot be due to residual methane absorption since this spectral region, located almost halfway between the CH$_4$ dyad and pentad regions, contains only hot-band lines whose intensities are negligible at Titan's temperatures. Furthermore, our methane linelist is based on variational calculations using ab initio potential and dipole moment surfaces \citep{Rey2018} that could not miss such a clear and localized absorption. Our model suffers from uncertainties in the methane far wing lineshape but this source of opacity, slowly varying with wavelength, cannot produce such a narrow absorption. We also checked whether any molecule detected in Titan's atmosphere exhibited an absorption band centered near 5.11 $\mu$m. The only one we found is the $\nu_6$ parallel band of propadiene (CH$_2$=C=CH$_2$), centered at 5.104 $\mu$m \citep{Pliva1982}, but its double-hump structure, with prominent P- and R-branches and a very weak Q-branch, does not match the shape of the observed 5.11-$\mu$m absorption.
   
    In addition, Figure~\ref{Titan3} shows that it does not behave like the weak CO absorption lines. In this figure, we compare the center-of-disk NIRSpec spectrum to the average  spectrum of all 0.1''$\times$0.1'' spatial pixels centered between 0.865 and 1.135 Titan radii from the disk center.
  While the CO line depths are similar in the two spectra, the depth of the 5.11-$\mu$m absorption, relative to the continuum, is reduced by about half. This strongly suggests that this absorption originates from the surface rather than from an atmospheric gas, with the surface contribution to the emitted flux being reduced from the center to the limb, while that of the haze is enhanced. For the same reason, the 5.11-$\mu$m absorption does not originate from the main haze, as it would then be enhanced at the limb. On the other hand, it could in principle be due to a thin condensate layer near the surface, beneath the bulk of the main haze. 
Nevertheless, a strong argument in favor of a surface origin of this feature is that it is observed on Pluto at a similar depth, where the atmosphere is optically much thinner.

\begin{table*}[ht!]
\caption{Gaussian fits of the 5.11-$\mu$m absorption}   
\label{table_1}  
\centering     
\begin{tabular}{l c c c c c c}    
\hline\hline  
Spectra &  \multicolumn{2}{c}{Center position }& Line-to-continuum &  \multicolumn{2}{c}{FWHM\tablefootmark{c}} & Residuals \\    
             & $\mu$m & cm$^{-1}$ & ratio (\%) & $\mu$m & cm$^{-1}$ & rms (I/F) \\       
\hline                 
   Titan/NIRSpec\tablefootmark{a} & 5.1126$\pm$0.0003 & 1955.9$\pm$0.1 & 5.8$\pm$0.2 & 0.0241$\pm$0.0008 & 9.2$\pm$0.3  & 0.00019 \\    
   Titan/MIRI\tablefootmark{b} &  5.1125$\pm$0.0007 & 1956.0$\pm$0.3 & 7.5$\pm$0.6 & 0.0180$\pm$0.0018 & 6.9$\pm$0.7 & 0.00068\\
   Pluto/MIRI & 5.1128$\pm$0.0021& 1955.9$\pm$0.8  & 4.5$\pm$0.5 & 0.069$\pm$0.008 & 26$\pm$3 & 0.0102 \\
   
\hline                               
\end{tabular}
\tablefoot{
\tablefoottext{a}{trailing side;}
\tablefoottext{b}{leading side;}
\tablefoottext{c}{Full Width at Half Maximum}
}
\end{table*}

  Fitting the 5.11-$\mu$m feature in the NIRSpec and MIRI spectra with a Gaussian function superimposed on a second-degree polynomial over the range 5.04--5.19 $\mu$m (Fig.~\ref{Titan1}), we obtained the parameters given in Table~\ref{table_1} with their error bars (derived from the residuals of the fit). The NIRSpec data indicate that the band is centered at 5.1126$\pm$0.0003 $\mu$m (1955.9$\pm$0.1 cm$^{-1}$) and has a full width at half maximum (FWHM) of 0.0241$\pm$0.0008 $\mu$m (9.2$\pm$0.3 cm$^{-1}$).  While the MIRI data indicate a position that agrees within error bars, the derived FWHM is 25\% smaller. The difference, which is significant at the 3-$\sigma$ confidence level, could be real given that the NIRSpec and MIRI instruments did not observe the same hemisphere. However, given the relatively large noise in the MIRI data, we prefer to await an analysis based on a more extended set of JWST data to conclude. Note that the equivalent width of the feature (i.e. its area, calculated here as the integral of the Gaussian function) is the same within error bars in the NIRSpec and MIRI data presented here ($\sim$0.0015 $\mu$m). As mentioned above, the surface signal is a bit diluted by the haze contribution so that the actual surface absorption depth and equivalent width are some $\sim$20\% larger.

\subsection{Pluto}

\begin{figure}[ht!]
   \centering
   \includegraphics[width=\hsize]{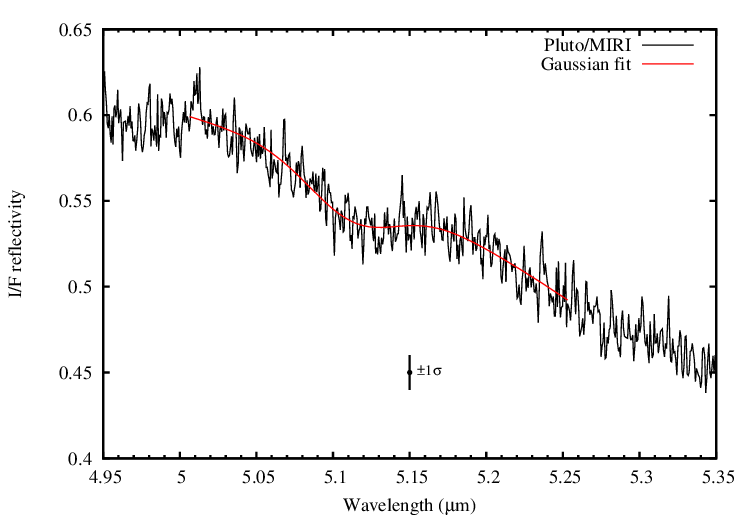}
      \caption{MIRI spectrum of Pluto from 4.95 to 5.35 $\mu$m (black). The red line represents a Gaussian fit of the absorption feature present at 5.11 $\mu$m using data points from 5.005 to 5.255 $\mu$m. The $\pm$1$\sigma$ error bar, based on the residuals of the fit, is indicated.} 
         \label{Pluto1}
   \end{figure}

Figure~\ref{Pluto1} shows a portion of the MIRI spectrum of Pluto presented in \cite{Lellouch2025}. The spectrum clearly shows a broad absorption centered at 5.1128$\pm$0.0021 $\mu$m (Table~\ref{table_1}), i.e at the same position as in the Titan spectra within error bars. On the other hand, the FWHM derived from the Gaussian fit (0.069$\pm$0.008 $\mu$m, i.e. 26$\pm$3 cm$^{-1}$), is $\sim$3 times larger than that in Titan's NIRSpec spectrum. The equivalent width ($\sim$0.0033 $\mu$m) is about twice that on Titan's surface.

\section{Discussion}
The 5.0-5.2 $\mu$m range is relatively poor in medium to strong bands of organic compounds \citep{Socrates2001}. We first checked that a 5.11-$\mu$m absorption peak does not appear in tholins produced in the laboratory by cold plasma discharge for any N$_2$--CH$_4$ mixture \citep{Brasse2015, Mathe2018, Drant2026}.  We then searched the literature for possible signatures of ice formed from species detected in Titan's atmosphere: \cite{Hudson2014b} for C$_2$H$_6$ and C$_2$H$_4$, \cite{Hudson2014a} and \cite{Abplanalp2019} for C$_2$H$_2$, \cite{Hudson2021} for C$_3$H$_8$, C$_3$H$_6$ and CH$_3$C$_2$H, \cite{Hudson2022a} for CH$_2$CCH$_2$, \cite{Schmitt2015} for C$_6$H$_6$, \cite{Moore2010} for HCN, C$_2$N$_2$, HC$_3$N, CH$_3$CN and C$_2$H$_5$CN,  \cite{DelloRusso1996} for C$_2$H$_3$CN and C$_4$N$_2$, and \cite{Bouilloud2015} for H$_2$O, CO$_2$ and CH$_4$. We did not find any band referenced in these publications that corresponds to the location of the observed absorption in Titan and Pluto. However, a signature may shift if the compound is mixed with other species. Below we examine three plausible candidates that exhibit an absorption band relatively close to 5.113 $\mu$m (1956 cm$^{-1}$).

The closest match is the weak $\nu_2$ acetylene (C$_2$H$_2$) band around 1961 cm$^{-1}$ (5.099 $\mu$m) \citep[][Hudson et al.\footnote{\url{https://science.gsfc.nasa.gov/691/cosmicice/constants.html}}]{Abplanalp2019}. This absorption is also seen in reflectance spectra of pure acetylene at $\sim$80 K ground to a fine powder \citep{Clark2010}. However, this spectrum also shows a stronger band centered at 4.83 $\mu$m causing a sharp decrease of the reflectance below 5.0 $\mu$m, at odds with the model variation of the surface albedo on Titan needed to reproduce the I/F NIRSpec spectrum (see above). Despite this possible inconsistency, we provisionally keep C$_2$H$_2$ ice as a potential candidate. Unpublished measurements made at IPAG (Grenoble) show that the signature of C$_2$H$_2$ ice diluted in N$_2$ with a 1\% concentration slightly shifts toward longer wavenumbers with respect to pure crystalline C$_2$H$_2$ (1964.2 vs.\ 1961.8 cm$^{-1}$, respectively), thus in the wrong direction. Since the effect of a polar matrix has not yet been tested, it is unclear whether this could shift the C$_2$H$_2$ ice signature by approximately -6 cm$^{-1}$ to the position of the feature observed by the JWST (1956 cm$^{-1}$).

Experiments attempting to identify the chemical alteration of ice mixtures due to cosmic rays demonstrate that irradiated H$_2$O:CH$_4$ ices lead to the formation of C$_2$H$_2$ with a spectral signature consistent with the observed feature at 1955 cm$^{-1}$ \citep{Mejia2020}. However, the $\sim$10 times stronger C$_2$H$_2$ ice band at 750 cm$^{-1}$  is not detected in the MIRI spectrum of Pluto \citep{Lellouch2025}, which does not give further support for this candidate on Pluto. More will be said below about the comparative effects of radiation on the two bodies.

The benzene molecule in the solid state exhibits a weak spectral signature close to 5.11~$\mu$m assigned to the $\nu_7$ + $\nu_{19}$ vibrational mode \citep{Nna-Mvondo22}. In the case of the pure crystalline phase at 130~K, this band shows two components at 1976 and 1981 cm$^{-1}$  (5.061 and 5.048  $\mu$m) \citep{Brown78, Schmitt2015}. In the amorphous or glassy phase (formed and measured at 15~K), the band displays a single component at 1966~cm$^{-1}$ (5.086 $\mu$m) \citep{Brown78, Nna-Mvondo22}. The width of this band is approximately 13~cm$^{-1}$ in the crystalline phase. In this spectral region, the $\nu_{11} + \nu_{19}$ mode is present at $\sim$1835~cm$^{-1}$ (5.45 $\mu$m) with a similar intensity, but is difficult to detect in Titan as it lies at the very edge of the atmospheric window. It is not detected on Pluto, and in this case an assignment  of the 5.11-$\mu$m absorption to C$_6$H$_6$ can be ruled out. The FWHM of the 5.11-$\mu$m band on Titan is 7-9~cm$^{-1}$, which is narrower than the above-mentioned experimental values. Overall, the presence of pure benzene appears precluded, but this does not rule out the presence of benzene mixed with other molecular species. Measurements by \cite{Brown78}, compared with earlier studies, show that the $\nu_7$ + $\nu_{19}$ mode exhibits a significant shift in position depending on its molecular environment: 1955, 1960, 1981, and 1972 cm$^{-1}$ when isolated in Ar, N$_2$, HCl, and Br$_2$ matrices, respectively. Unfortunately, no data are available to assess the corresponding FWHMs. In conclusion, benzene mixed with other molecular species remains a possible candidate in the case of Titan, but a definitive conclusion cannot be drawn without additional laboratory experiments.

The group of allenes (organic compounds that include a C=C=C pattern) show a C=C=C out-of-phase stretching vibration mode in the 1900--2000 cm$^{-1}$ range \citep{LinVien1991}. This group is essentially the only one among organic compounds to exhibit strong absorption bands in this range \citep{Socrates2001}. The exact position depends on the nature of the molecular groups attached to either end of the carbon chain. The simplest allene, propadiene, detected in Titan's atmosphere \citep{Lombardo2019}, exhibits strong signatures at 1947.3 and 1948 cm$^{-1}$ (5.135 and 5.133 $\mu$m), for the crystalline (80~K) and amorphous (8~K) phases, respectively \citep{Hudson2022a}. At first glance, this compound appears as a plausible candidate because it shows only one strong feature in the 5.0-5.2 $\mu$m range, and its width of 5.2 cm$^{-1}$ is smaller than the width of the observed 5.11-$\mu$m  (1956 cm$^{-1}$) feature. The broadening could then be interpreted as the result of physical effects like grain size. Nevertheless, the spectral shift of $\sim$10 cm$^{-1}$ shows that pure propadiene is not a good candidate. Systematic laboratory measurements are necessary at that point to investigate the role of the chain length, cross-linking and nature of the branching species on the width, shape and position of the band of allenes. 

Last, two candidates with weak features around 5.113 $\mu$m should be mentioned. The first is ketene (CH$_2$C=O), which exhibits in the solid state a weak spectral signature ($2\nu_6$) at 1942 and 1947 cm$^{-1}$ (5.149 and 5.136 $\mu$m), in the pure phase and when isolated in an argon matrix, respectively \citep{Moore63}. This feature is accompanied by a much more intense band at $\sim$2080 cm$^{-1}$ (4.81 $\mu$m) usually assigned to out-of-phase stretching modes. The peak positions 1942 and 1947 cm$^{-1}$ are not that far from 1956 cm$^{-1}$ but, unfortunately, no studies provide a systematic analysis of the parameters controlling the position and width of this feature and no further conclusion can be drawn. As for the second candidate, a weak band at 1957~cm$^{-1}$ (5.110 $\mu$m) has been identified in the irradiation residue of methanol ice \citep{Quirico2023}. This band may be attributed either to an allene-type compound or to a C=C=O-type functional group. A radiolytic origin appears unlikely in the case of Titan but remains plausible for Pluto.

In addition, HCN deserves some consideration because hydrogen bonds may a priori induce substantial shifts of the band frequencies when the molecule is diluted in another ice. HCN exhibits a band ($\nu_3$) at 4.76 $\mu$m (2100 cm$^{-1}$), which is primarily controlled by the CN stretch. The peak position of this band varies between 2100 and 2100.5 cm$^{-1}$ over the temperature range 25-120 K for the type I tetragonal crystalline phase, and over 50-110 K for the amorphous phase, respectively \citep{DelloRusso1996, Moore2010}. These values are close to those of the molecule in the gas phase or isolated in an Ar matrix, indicating that the hydrogen bond that links H and N atoms is too weak to induce a significant spectral shift \citep{Muller93}. Experiments conducted on HCN diluted into a variety of polar and apolar matrices show peak positions in the range 2100.3-2063.5 cm$^{-1}$, which are too far from the position of the 5.11-$\mu$m band \citep{Ozhiganov24}; therefore, we do not consider HCN to be a candidate for the 5.11-$\mu$m absorption.

A significant finding of our study is that the width of the 5.11-$\mu$m feature is $\sim$3 times larger on Pluto than on Titan. If the same molecular species is responsible for this signature on both bodies, several mechanisms can be considered to explain the larger width observed on Pluto. First, grain size affects the band depth as this parameter primarily controls the optical path length. 
More pronounced broadening effects can be observed in the case of saturated bands, for which the band wings are considerably amplified. However, the 5.11-$\mu$m absorption has low intensity,  its Gaussian profile is not consistent with spectral saturation, and these effects are not expected. Similarly, macroscopic mixing regimes (intimate vs.\ areal) do not lead to very pronounced broadening effects.

Pluto's surface temperature (30-60 K) is lower than that of Titan (90-95 K). In the case of ices, the general trend is for bands to broaden as temperature increases, which is inconsistent with the observations. A more likely mechanism is related to the physical state of the molecular species involved, and more specifically to the diversity of its environment at the molecular scale.  Concentration-dependent effects may arise when molecular clusters of different sizes are formed \citep{Quirico1997, Behringer1958}. Single, double, triple and larger clusters each exhibit absorption bands with slightly different position and width, resulting in broadening of the overall spectral feature. In the case of organic compounds such as the allenes mentioned earlier, the nature of the chemical groups R$_1$ and R$_2$ attached to the terminal carbon atoms of the C=C=C moiety strongly influences the position and width of the absorption band. Consequently, a complex mixture of organic molecules exhibiting a wide diversity of R$_1$ and R$_2$ groups may generate substantial broadening.

The surfaces of Titan and Pluto are exposed to energetic ions for periods long enough to trigger radiolytic and chemical processes. In the case of Titan, the charged particles are primarily secondary electrons produced during interactions with galactic cosmic rays (GCRs) near 65 km \citep{Gronoff2011}, whereas Pluto's surface is irradiated by GCR ions that cover a broad range of energies. Since Pluto's atmosphere is thinner, a fraction of the GCR penetrates the surface to a depth ranging from several centimeters to several tens of centimeters. 
The effects of irradiation are manifold: radiolysis (breaking of chemical bonds), sputtering (e.g., H$_2$), formation of new species through radical recombination, and structural transformation (e.g., amorphization). All these processes can lead to a diversification of the molecular environment, the broadening of a spectral signature, or a change in its spectral shape. 
The low-energy population of the GCR ions (except hydrogen) has a substantial nuclear stopping power and is more effective at destabilizing carbonaceous compounds \citep{Faure2021}. However, these processes depend both on the molecular species and on the environment with which they are likely to interact.

\section{Conclusion}

We have found an unidentified absorption feature in JWST spectra of Titan (trailing and leading sides) and of Pluto. This absorption is centered at 5.113 $\mu$m (1956~cm$^{-1}$) and very likely originates from the surface of the two bodies. Its width is about three times larger on Pluto than on Titan's trailing side. On Titan, it might be narrower on the leading hemisphere than on the trailing hemisphere. We searched for possible simple ice and organic candidates in the literature and found that only few of them are plausible: the group of allenes and, if mixed with other species, benzene, ketene or less likely acetylene (on Titan). The difference in the width of the feature between the two bodies is most likely due to the physical state of the unknown compound at the molecular scale.

In the near future, mapping the characteristics of this absorption feature over Titan's disk using a more complete JWST data set may help to understand its nature and origin. Dragonfly, the next mission to Titan scheduled to arrive in the mid-2030s, will conduct in situ studies of the surface composition and chemical complexity at various geological sites. In particular, the Dragonfly Mass Spectrometer (DraMS) should be able to identify some of the potential candidates for the 5.11-$\mu$m absorption, although the relatively volatile acetylene may not be retained during analysis (M.\ Trainer, pers. comm., 2026). However, the lack of onboard infrared spectroscopy capabilities prevents any direct observation of the spectral feature itself in surface materials.

\begin{acknowledgements}
      This work is based on observations made with the NASA/ESA/CSA James Webb Space Telescope. The data were obtained from the Mikulski Archive for Space Telescopes at the Space Telescope Science Institute, which is operated by the Association of Universities for Research in Astronomy, Inc., under NASA contract NAS 5-03127 for JWST. These observations are associated with programs 1251 and 1658. BB, EL and SR acknowledge support from the Programme National de Plan\'etologie (PNP) of CNRS-INSU co-funded by CNES. SR further acknowledges support from CNES and the Agence Nationale de la Recherche (ANR) through the project RaD3-net (ANR-21-CE49-0020). Part of this work was carried out at the Jet Propulsion Laboratory, California Institute of Technology, under a contract with the National Aeronautics and Space Administration. MC acknowledges support from the Caltech Discovery Fund. CAN received support for the work from NASA GSFC Strategic Science Fund. NT is supported by UK Science and Technology Facilities Council grant ST/Y000676/1. N.P.-A. acknowledges the Ministry of Science, Innovation, and Universities (MCIU) in Spain and the State Agency for Research (AEI) for funding through the ATRAE programme, project ATR2023-145683. BB thanks S.\ Vinatier for helpful discussions.
\end{acknowledgements}

 \bibliographystyle{aa} 
 \bibliography{aa60810-26}

\end{document}